\documentclass[a4paper,12pt]{article}

\pdfoutput=1

\usepackage[tbtags]{amsmath}
\usepackage{amssymb}
\usepackage{ifthen}
\usepackage{slashed}
\usepackage{amsfonts}
\usepackage{mathrsfs}
\usepackage{bm}
\usepackage{graphicx,subfigure,booktabs}
\usepackage[numbers,sort&compress]{natbib}
\usepackage{verbatim}
\usepackage{appendix}

\usepackage[colorlinks,
            linkcolor=black,
            filecolor=black,
            anchorcolor=black,
            urlcolor=black,
            citecolor=blue
            ]{hyperref}

\usepackage{color}
\usepackage{ulem}
\usepackage{setspace}
\numberwithin{equation}{section}

\newlength{\dinwidth}
\newlength{\dinmargin}
\makeatletter
\newcommand{\thickhline}{%
    \noalign {\ifnum 0=`}\fi \hrule height 1pt
    \futurelet \reserved@a \@xhline
}
\makeatother

\allowdisplaybreaks

\begin{document}

\title{}

\title{\bf Heavy quarkonium decay $V \rightarrow ggg$ with both relativistic and QCD radiative corrections}

\author{Hong-Mei Jiang$^{a}$, Chao-Jie Fan$^{a,b}$\footnote{fancj@hbnu.edu.cn}, Jun-Kang He$^{a,b}$\footnote{hejk@hbnu.edu.cn}, and Cui Kong$^{a}$\footnote{cuikong@hbnu.edu.cn}\\[15pt]
{$^a$\small College of Physics and Electronic Science, Hubei Normal University, Huangshi 435002, China}\\[0.2cm]
{$^b$\small Key Laboratory of Quark and Lepton Physics (MOE), }\\
{\small Central China Normal University, Wuhan 430079, China}}
\date{}


\maketitle
\vspace{0.2cm}

\begin{abstract}
{\noindent} In the heavy quarkonium decay process $V \to ggg$ ($V=J/\psi, \Upsilon$), making a definite prediction including relativistic corrections has so far remained a significant challenge. In this work, we study this decay process by taking into account the relativistic corrections in the Bethe-Salpeter formalism, where the relativistic bound-state wave function of quarkonium is obtained by solving the Bethe-Salpeter equation under the covariant instantaneous ansatz. Through analytical calculation, we find that some polarized decay widths vanish due to the helicity selection rule, which suppresses the corresponding helicity amplitudes. Owing to helicity flip symmetry and phase space symmetry, the nonvanishing polarized decay widths are not all independent; they are related through a set of symmetry relations. Then we obtain the unpolarized decay width formula $\Gamma(V \rightarrow ggg)=\frac{80(\pi^{2}-9)\alpha_{s}^{3}N_{V}^{2}\beta_{V}^{3}}{81\pi^{9/2} M } (1-\kappa\frac{\beta_{V}^{2}}{M^{2}})$, where the factor $\kappa\frac{\beta_{V}^{2}}{M^{2}}$ arises from the relativistic corrections with $\kappa\equiv\frac{3(112+25\pi^{2})}{16(\pi^{2}-9)}$. Furthermore, including both relativistic and QCD radiative corrections within the factorization assumption, our predictions of $\mathcal{B}(V \to ggg)$ and $\mathcal{B}(V \to e^{+}e^{-})$ agree well with their experimental data. As a crossing check, with the experimental value of the ratio $R_{V} = \frac{\Gamma(V \to ggg)}{\Gamma(V \to e^{+}e^{-})}$ and our result for $R_{V}$, we extract $\alpha_{s}(M_{J/\psi}/2)=0.31$ and $\alpha_{s}(M_{\Upsilon}/2)=0.20$, respectively.
\end{abstract}


\newpage

\section{INTRODUCTION}
\label{sec:intro}

Since the discovery of the charmonium $J/\psi$~\cite{E598:1974sol, SLAC-SP-017:1974ind}, remarkable experimental progress in heavy quarkonium physics has been achieved at facilities including LHCb, Belle, BaBar, CDF, BESIII, and CLEO-c~\cite{Brambilla:2010cs, BaBar:2013byz, Belle-II:2018jsg, BESIII:2020nme, Zhao:2025sna}. This field has also attracted significant theoretical attention because heavy quarkonium systems provide an ideal testing ground for understanding the interplay between perturbative and nonperturbative regimes of quantum chromodynamics (QCD)~\cite{Zhang:2008gp, Fan:2009zq, Wang:2011qg, Xu:2012am, Dong:2012xx, He:2019mpy, Fan:2019sap, He:2020jvj}. For the heavy quarkonium $J/\psi$ ($\Upsilon$) with the quantum numbers $J^{PC}=1^{--}$~\cite{ParticleDataGroup:2024cfk}, its various decay channels have been the essential subject of many experimental and theoretical studies. Among these channels, the three-gluon decay of the vector quarkonium $J/\psi$ is particularly significant, since it provides us with invaluable information on the strong interactions between quarks and gluons in intermediate energy region, where the interplay between perturbative and nonperturbative effects of QCD becomes important.

The heavy quarkonium $V$ ($V=J/\psi, \Upsilon$) cannot decay directly into a single gluon due to the color confinement. Furthermore, its decay into two gluons is forbidden by $C$-parity conservation. Therefore, the gluonic decay of $V$ is the three-gluon decay $V \rightarrow ggg$. Theoretically, this decay provides a clean environment to probe the intrinsic structure of heavy quarkonium, since it does not involve the hadronization of the final-state gluons. To lowest order in QCD~\cite{Appelquist:1974zd, Appelquist:1975ya}, the decay width, which is proportional to the modulus squared of the radial wave function at origin $|R_{V}(0)|^{2}$, can be obtained by supplementing the three-photon decay width of ortho-positronium~\cite{Ore:1949te} with the appropriate color factor. The ratio of the gluonic to leptonic widths of $V$ is predicted to be~\cite{Appelquist:1974zd, Appelquist:1975ya} $R_{V}=\frac{\Gamma(V \rightarrow ggg)}{\Gamma(V \rightarrow e^{+}e^{-})}=\frac{10(\pi^{2}-9)\alpha_{s}^{3}}{81\pi e_{Q}^{2}\alpha^{2}}$ with $e_{Q}$ the charge of the $Q$ ($Q=c, b$) quark, which is independent of the modulus squared $|R_{V}(0)|^{2}$. Since this ratio exhibits strong sensitivity only to the QCD running coupling constant $\alpha_{s}$, it provides a powerful method for estimating $\alpha_{s}$. By using the experimental value of this ratio, one obtains $\alpha_{s}(m_{c})=0.195$ and $\alpha_{s}(m_{b})=0.181$, which are substantially lower than the values obtained from other decays~\cite{Bethke:2006ac}. This may imply that the higher-order effects play an important role in the decay process $V \rightarrow ggg$.

The first calculation of the QCD radiative corrections to the decay $\Upsilon \rightarrow ggg$ was performed numerically by Mackenzie and Lepage~\cite{Mackenzie:1981sf}. Including the first-order QCD radiative corrections, the ratio $R_{V}$ is predicted to be $R_{V} = \frac{10(\pi^{2}-9)\alpha_{s}^{3}} {81\pi e_{Q}^{2}\alpha^{2}} (1+C\frac{\alpha_{s}}{\pi})$ with $C=1.63$ for $J/\psi$, $0.43$ for $\Upsilon$~\cite{Kwong:1987ak}. With the improved prediction and the experimental value of the ratio, one extracts $\alpha_{s}(m_{c})=0.189$ and $\alpha_{s}(m_{b})=0.180$, which are even slightly lower than the leading-order estimates. Recently, calculations of the decay width $\Gamma(V \rightarrow ggg)$ based on a potential model~\cite{Kher:2018wtv, Soni:2020tji} and a nonrelativistic quark model~\cite{Segovia:2016xqb, Mistry:2024lco} have been updated to include first-order QCD radiative corrections. However, these updates have not yielded a substantial improvement. This suggests that the relativistic corrections play a key role in these decay processes.

The first-order relativistic corrections to the decay $V \rightarrow ggg$ were first calculated by Keung and Muzinich~\cite{Keung:1982jb}. In their work, only the kinematic relativistic corrections were included. Later, Chao et al.~\cite{Chao:1995cz} arrived at a result compatible with experiment by including both kinematic and dynamical relativistic corrections, albeit with an ad hoc approximation employed for the hard-scattering kernel ($\int \mathrm{d}^{3}q f(\mathbf{q}) (1-2.70 \frac{\mathbf{q}^{2}}{m^{2}})\approx \int \mathrm{d}^{3}q f(\mathbf{q}) (1+2.70 \frac{\mathbf{q}^{2}} {m^{2}})^{-1}$). Subsequently, the ${\cal O}(v^{4})$ corrections were calculated within nonrelativistic QCD (NRQCD)~\cite{Bodwin:2002cfe, Brambilla:2008zg}. However, the predictive power of this approach is limited by the proliferation of poorly constrained nonperturbative matrix elements, especially for the higher-order corrections. Recently, Sang et al.~\cite{Sang:2020zdv} computed the ${\cal O}(\alpha_{s}v^{2})$ corrections within NRQCD. They found that these corrections failed to counterbalance the large negative corrections from the ${\cal O}(v^{2})$ terms and thus still yielded an unphysical negative decay width for $J/\psi$ channel. Apparently, although relativistic corrections play a crucial role~\cite{Eichten:2007qx}, current methods for handling them remain somewhat unsatisfactory. Therefore, it is worth reexamining these corrections in detail and clarifying the full dynamical mechanisms underlying the decay process $V \rightarrow ggg$.

In addition to the aforementioned methods, the decay process $J/\psi (\Upsilon) \rightarrow ggg$ has also been investigated using phenomenological models which introduce an effective gluon mass~\cite{Consoli:1993ew, Consoli:1997ts, Mihara:2000wf}. Generally, these studies could obtain $\alpha_{s} (m_{c})\sim 0.3$ and $\alpha_{s} (m_{b})\sim 0.2$ with a large effective gluon mass ($M_{g} = 0.6 \sim 1.2\, \mathrm{GeV}$). However, this assumption may require further physical justification.

In this paper, to clarify the dynamical picture of the decay $J/\psi (\Upsilon) \rightarrow ggg$, we employ the Bethe-Salpeter (B-S) framework ~\cite{Salpeter:1951sz, Salpeter:1952ib, Mitra:1990av, Bhatnagar:2009jg, Bhatnagar:2013bha, Bhatnagar:2016otj, Gebrehana:2019mpw, He:2020kin} to analytically calculate the wave functions of the heavy quarkonia $J/\psi$ and $\Upsilon$, as well as their decay amplitudes. A key aspect of our approach, which enables a detailed reexamination of the relativistic corrections, is that the internal momentum of the quarkonium is retained in both the soft bound-state wave function and the hard-scattering amplitude. This treatment allows the framework to naturally handle the relativistic corrections, which include the dynamical corrections from the bound-state wave function and the kinematical corrections from the hard-scattering amplitude. With a QCD-motivated B-S equation under the covariant instantaneous ansatz (CIA)~\cite{Mitra:1990av, Bhatnagar:2009jg, Bhatnagar:2013bha, Bhatnagar:2016otj, Gebrehana:2019mpw}, we obtain the analytic forms of wave functions by employing the full B-S kernel, which includes both the long-range confinement and the short-range one-gluon-exchange (Coulomb) interactions. These wave functions are then used to calculate the decay width $\Gamma (J/\psi (\Upsilon) \rightarrow ggg)$. Owing to helicity flip and phase space symmetries, we find that the polarized decay widths ($\Gamma_{[\lambda_V,\lambda_1,\lambda_2,\lambda_3]}^{ggg}$) can be divided into four groups. Furthermore, based on the factorization assumption, we take into account both relativistic and QCD radiative corrections to next-to-leading order. And our theoretical predictions for the branching ratios $\mathcal{B}(V \rightarrow ggg)$ and $\mathcal{B}(V \rightarrow e^{+}e^{-})$ are consistent with experimental data. In addition, the QCD running coupling constant $\alpha_{s}$ is extracted from our calculations and experimental measurements of the gluonic and leptonic decay widths.

The paper is organized as follows. The theoretical framework and the analytical calculation for the decay process $J/\psi (\Upsilon) \rightarrow ggg$ is presented in detail in Sec.~\ref{sec:framework}. In Sec.~\ref{sec:Results and discussions} we show our results and phenomenological discussions. The last section is our summary.

\section{THEORETICAL FRAMEWORK}
\label{sec:framework}
\subsection{Bethe-Salpeter equation}
\label{subsec:B-S equation}

It is generally known that the B-S equation~\cite{Salpeter:1951sz,Salpeter:1952ib} is an effective relativistic equation with a solid basis in quantum field theory, making it a conventional approach for treating various bound-state problems. In this subsection, we briefly review the formulation of the B-S framework. For heavy quarkonium $V$ ($J/\psi$ or $\Upsilon$),
the B-S equation has the form~\cite{Mengesha:2011pu, Negash:2015hma, Negash:2015rua, Bhatnagar:2016otj, Gebrehana:2019mpw, He:2020kin, Fan:2024lef}
\begin{equation}\label{bse}
S^{-1}_{F}(f)\Psi(K,q)S^{-1}_{F}(-\bar{f}) =
\int\frac{\mathrm{d}^{4}q^{\prime}}{(2\pi)^{4}}\Big{[}-i\mathcal{K}(K,q,q^{\prime})\Psi(K,q^{\prime})\Big{]},
\end{equation}
where $\mathcal{K}(K,q,q^{\prime})$ represents the interaction kernel between the internal quark and antiquark, and $S_{F}(p)=i/(\slashed{p}-\hat{m}+i\epsilon)$ represents the propagator with
the effective mass of $c$ ($b$) quark $\hat{m}$. The momenta of the quark and antiquark can be written as
\begin{equation}
f=\frac{K}{2}+q,~~~~\bar{f}=\frac{K}{2}-q,
\end{equation}
where $q$ and $K$ represent the internal momentum and the total momentum of the heavy quarkonium, respectively.

For convenience,  the internal momentum $q$ is decomposed into a transverse component $\hat{q}$ (with $\hat{q} \cdot K = 0$) and a longitudinal component $q_{\parallel}$ (parallel to the total momentum $K$):
\begin{eqnarray}
q^{\mu}&=& q_{\parallel}^{\mu}+\hat{q}^{\mu}, \nonumber\\
q_{\parallel}^{\mu} &=& \frac{q_{K}}{M} K^{\mu}.
\end{eqnarray}
Here both $q_{K}=q\cdot K / M$ and $\hat{q}^{2}=q^{2}-q_{K}^{2}$ are Lorentz invariant variables and $M$ is the mass of the heavy  quarkonium.

The CIA~\cite{Mitra:1990av, Bhatnagar:2009jg, Bhatnagar:2013bha, Bhatnagar:2016otj, Gebrehana:2019mpw} posits that the interaction is instantaneous. Consequently, the interaction kernel $\mathcal{K}(K,q,q^{\prime})$ is independent on the timelike component of the internal momentum:
\begin{equation}\label{k}
\mathcal{K}(K,q,q^{\prime})=V(\hat{q},\hat{q}^{\prime}).
\end{equation}
In this framework, the interaction kernel $V(\hat{q},\hat{q}^{\prime})$ typically includes both long-range confinement and one-gluon exchange interactions~\cite{Bhatnagar:2016otj, Gebrehana:2019mpw}.

Then the B-S wave function can be expressed as
\begin{equation}\label{bsecia}
\Psi(K,q)=-S_{F}(f)\Gamma(\hat{q})S_{F}(-\bar{f}),
\end{equation}
where the hadron-quark vertex function reads
\begin{equation}
   \Gamma(\hat{q})=i\int\frac{\mathrm{d}^{4}q^{\prime}}
   {(2\pi)^{4}}V(\hat{q},\hat{q}^{\prime})\Psi(K,q^{\prime})=\int\widetilde{\mathrm{d}^{3}q^{\prime}}
   V(\hat{q},\hat{q}^{\prime})\psi(\hat{q}^{\prime})
\end{equation}
with the Salpeter wave function $\psi(\hat{q})=\frac{i}{2\pi}\int\mathrm{d}q_{K}\Psi(K,q)$ and $\widetilde{\mathrm{d}^{3}q^{\prime}}=\mathrm{d}^{3}q^{\prime}/(2\pi)^{3}$. We now introduce the projection operator~\cite{Negash:2015hma,Negash:2015rua,Bhatnagar:2016otj,Gebrehana:2019mpw}
\begin{equation}
  \Lambda^{\pm}_{i}(\hat{q})=\frac{1}{2\omega}\left[\frac{\slashed{K}}{M}\omega
  \pm(-1)^{(i+1)}(\hat{m}+\slashed{\hat{q}})\right],
\end{equation}
which allows us to decompose the propagators as follows:
\begin{eqnarray}
\frac{1}{\slashed{f}-\hat{m}+i\epsilon }&=&\frac{\Lambda^{+}_{1}(\hat{q})}{q_{K}+\frac{M}{2}-\omega+i\epsilon}
  +\frac{\Lambda^{-}_{1}(\hat{q})}{q_{K}+\frac{M}{2}+\omega-i\epsilon},\nonumber\\
 \frac{1}{\slashed{\bar{f}}+\hat{m}-i\epsilon}&=& \frac{\Lambda^{+}_{2}(\hat{q})}{-q_{K}+\frac{M}{2}-\omega+i\epsilon}
  +\frac{\Lambda^{-}_{2}(\hat{q})}{-q_{K}+\frac{M}{2}+\omega-i\epsilon}
\end{eqnarray}
with $\omega=\sqrt{\hat{m}^{2}-\hat{q}^{2}}$. Performing the $q_{K}$-integration of Eq.~\eqref{bsecia},
one can obtain~\cite{Negash:2015hma,Negash:2015rua,Bhatnagar:2016otj,Gebrehana:2019mpw}
\begin{eqnarray}\label{Salpeterequations}
(M-2\omega)\psi^{++}(\hat{q})&=&-\Lambda^{+}_{1}(\hat{q})\Gamma(\hat{q})\Lambda^{+}_{2}(\hat{q}),\nonumber \\
(M+2\omega)\psi^{--}(\hat{q})&=&\Lambda^{-}_{1}(\hat{q})\Gamma(\hat{q})\Lambda^{-}_{2}(\hat{q}),\nonumber \\
\psi^{+-}(\hat{q})&=&0,\nonumber\\
\psi^{-+}(\hat{q})&=&0
\end{eqnarray}
with $\psi^{\pm\pm}(\hat{q})=\Lambda^{\pm}_{1}(\hat{q})\frac{\slashed{P}}{M}\psi(\hat{q})
\frac{\slashed{P}}{M}\Lambda^{\pm}_{2}(\hat{q})$ and
$\psi(\hat{q})=\psi^{++}(\hat{q})+\psi^{+-}(\hat{q})+\psi^{-+}(\hat{q})+\psi^{--}(\hat{q})$.

For the quarkonium $V$ with quantum numbers $J^{PC}=1^{--}$, the general form of Dirac structures of the wave function up to sub-leading order can be expressed as~\cite{LlewellynSmith:1969az, Wang:2005qx, Bhatnagar:2013bha, Negash:2015rua}
\begin{eqnarray}\label{psihatq}
\psi(\hat{q})&=&M\slashed{\varepsilon} f_{1}(\hat{q})+\slashed{\varepsilon}\slashed{K} f_{2}(\hat{q})
+[\slashed{\varepsilon}\slashed{\hat{q}}-\hat{q}\cdot \varepsilon]f_{3}(\hat{q})\nonumber\\
& &+[\slashed{K}\slashed{\varepsilon}\slashed{\hat{q}}-(\hat{q}\cdot\varepsilon)\slashed{K}]
\frac{f_{4}(\hat{q})}{M}+(\hat{q}\cdot\varepsilon)f_{5}(\hat{q})+(\hat{q}\cdot\varepsilon)
\slashed{K}\frac{f_{6}(\hat{q})}{M},
\end{eqnarray}
where scalar function $f_{i}(\hat{q})$ ($i=1,2,\cdots,6$) depends on $\hat{q}^{2}$, and $\varepsilon$ is the polarization vector of $V$. Since the contributions associated with the $\hat{q}^{2}$-order Dirac structures are suppressed~\cite{Bhatnagar:2013bha}, we are interested in calculations up to the $\hat{q}$-order terms in this work. The two homogeneous equations in~\eqref{Salpeterequations}, namely $\psi^{+-}(\hat{q}) = 0$ and $\psi^{-+}(\hat{q}) = 0$, act as two independent constraint conditions. Substituting the general form of $\psi(\hat{q})$ into these constraints and requiring the coefficients of the independent Dirac structures to vanish, we obtain the following relations among the scalar functions:
\begin{eqnarray}
f_{3}(\hat{q})=f_{6}(\hat{q})=0,\quad\quad  f_{4}(\hat{q})=-\frac{M}{\hat{m}}f_{2}(\hat{q}),\quad\quad f_{5}(\hat{q})=\frac{M}{\hat{m}}f_{1}(\hat{q}).
\end{eqnarray}
As a result, the Salpeter wave function simplifies to
\begin{eqnarray}
\psi(\hat{q})&=&(M\slashed{\varepsilon} + \frac{M}{\hat{m}} (\hat{q}\cdot \varepsilon))f_{1}(\hat{q}) + (\slashed{\varepsilon}\slashed{K} + \frac{\slashed{K}(\hat{q}\cdot \varepsilon)}{\hat{m}} - \frac{\slashed{K}\slashed{\varepsilon}\slashed{\hat{q}} }{\hat{m}}) f_{2}(\hat{q}).
\end{eqnarray}
Next, employing the local approximation
\begin{equation}
V(\hat{q},\hat{q}^{\prime})=(2\pi)^{3}\delta^{3}(\hat{q}-\hat{q}^{\prime})
\bar{V}(\hat{q}^{\prime})\gamma^{\mu}\otimes\gamma_{\mu}
\end{equation}
and the approximate relation
\begin{equation}
\gamma^{\mu}\psi(\hat{q})\gamma_{\mu} \approx \Theta \psi(\hat{q})
\end{equation}
with $\Theta = -2$ for vector meson~\cite{Negash:2015rua}, and applying the first two Salpeter equations in~\eqref{Salpeterequations}, we obtain two identical decoupled equations
\begin{eqnarray}
\left(\frac{M^{2}}{4}-\hat{m}^{2}+\hat{q}^{2}\right)f_{1}(\hat{q})&=&\hat{m} \Theta \bar{V}(\hat{q}) f_{1}(\hat{q}),\nonumber\\
\left(\frac{M^{2}}{4}-\hat{m}^{2}+\hat{q}^{2}\right)f_{2}(\hat{q})&=&\hat{m} \Theta \bar{V}(\hat{q}) f_{2}(\hat{q}).
\end{eqnarray}
Since $f_1(\hat{q})$ and $f_2(\hat{q})$ satisfy the same differential equation, they are proportional to each other. The constant of proportionality can be determined by the structure of the wave function in the non-relativistic limit, i.e., $(M\slashed{\varepsilon}-\slashed{\varepsilon}\slashed{K})f(\hat{q})$. Then we further obtain
\begin{eqnarray}
f_{1}(\hat{q})=-f_{2}(\hat{q})=f(\hat{q}),
\end{eqnarray}
and the Salpeter wave function
\begin{eqnarray}
\psi(\hat{q})&=& \left( M\slashed{\varepsilon} - \slashed{\varepsilon}\slashed{K} + \frac{M}{\hat{m}} (\hat{q}\cdot \varepsilon) - \frac{\slashed{K}}{\hat{m}}(\hat{q}\cdot \varepsilon) + \frac{\slashed{K}\slashed{\varepsilon}\slashed{\hat{q}} }{\hat{m}} \right) f(\hat{q}).
\end{eqnarray}
In the rest frame of the heavy quarkonium $V$, the momenta $K$ and $\hat{q}$ are given by
\begin{eqnarray}
K^{\mu}&=&(M,\mathbf{0}),\quad\quad   \hat{q}^{\mu}=(0,\mathbf{\hat{q}})=(0,\mathbf{q}).
\end{eqnarray}
Adopting the approximation $\hat{m} \approx M/2$ and a potential $\bar{V}(\hat{q})$ that includes both confinement and one-gluon exchange interactions~\cite{Bhatnagar:2016otj, Gebrehana:2019mpw}, the wave function can be reduced to
\begin{eqnarray}
\psi(\hat{q})&=& \left( M\slashed{\varepsilon} - \slashed{\varepsilon}\slashed{K} + 2 (\hat{q}\cdot \varepsilon) - \frac{2 \slashed{K}(\hat{q}\cdot \varepsilon)}{M} + \frac{2 \slashed{K}\slashed{\varepsilon}\slashed{\hat{q}} }{M} \right) f(\hat{q})
\end{eqnarray}
with the scalar function
\begin{eqnarray}\label{fhatq}
f(\hat{q})&=&N_{V}\frac{1}{\pi^{3/4}}\frac{1}{\beta^{3/2}_{{V}}} e^{-\frac{\mathbf{\hat{q}}^{2}}{2\beta^{2}_{V}}},
\end{eqnarray}
where $N_{V}$ is a constant and $\beta_{V}$ is the harmonic oscillator parameter. Following the procedure established in Refs.~\cite{Bhatnagar:2016otj, Gebrehana:2019mpw}, we take $N_{J/\psi} = 0.57 \, \mathrm{MeV}^{-1/2}$, $\beta_{J/\psi} = 310 \, \mathrm{MeV}$ for the $J/\psi$; $N_{\Upsilon} = 0.30 \, \mathrm{MeV}^{-1/2}$, $\beta_{\Upsilon} = 600 \, \mathrm{MeV}$ for the $\Upsilon$.
For convenience, we employ the shorthand
\begin{eqnarray}
\psi(\hat{q})&=& f(\hat{q}) {\cal P}(\hat{q}),
\end{eqnarray}
where ${\cal P}(\hat{q})$ denotes the Dirac structure
\begin{eqnarray}
{\cal P}(\hat{q})&=& M\slashed{\varepsilon} - \slashed{\varepsilon}\slashed{K} + 2 (\hat{q}\cdot \varepsilon) - \frac{2 \slashed{K}(\hat{q}\cdot \varepsilon)}{M} + \frac{2 \slashed{K}\slashed{\varepsilon}\slashed{\hat{q}} }{M}.
\end{eqnarray}

\subsection{The decay $V \rightarrow ggg$ in perturbative QCD}
\label{subsec:V-ggg}

It is well known that the decay amplitude of $V \rightarrow ggg$ can be factorized into a short-distance part, which can be calculated using perturbative methods, and a long-distance part, which describes the nonperturbative dynamics of the bound state. In the rest frame of the heavy quarkonium $V$, the decay amplitude of $V \rightarrow ggg$ can be written as
\begin{eqnarray}
{\mathcal A}&=&\sqrt{N_{c}}\int\frac{\mathrm{d}^{4}q}{(2\pi)^{4}}\textrm{Tr}\left[\Psi(K,q){\cal O}(f,\bar{f})\right],
\end{eqnarray}
where the factor $\sqrt{N_{c}}$ accounts for the color properties of the quark-antiquark content, ${\cal O}(f,\bar{f})$ is the hard-scattering amplitude, and the momenta of the quark and antiquark are given by
\begin{eqnarray}
f^{\mu}&=&\frac{K^{\mu}}{2}+q^{\mu}
=\left(\frac{M}{2}+q^{0},\mathbf{q}\right),\nonumber\\
\bar{f}^{\mu}&=&\frac{K^{\mu}}{2}-q^{\mu}
=\left(\frac{M}{2}-q^{0},-\mathbf{q}\right).
\end{eqnarray}
Under the CIA, a more relevant treatment is to take $q^{0}\ll M$, so one can obtain the momenta
\begin{eqnarray}
f^{\mu}&\approx&\left(\frac{M}{2},\mathbf{q}\right)=\frac{K^{\mu}}{2}+\hat{q}^{\mu},\nonumber\\
\bar{f}^{\mu}&\approx&\left(\frac{M}{2},-\mathbf{q}\right)=\frac{K^{\mu}}{2}-\hat{q}^{\mu},
\end{eqnarray}
and a simplified hard-scattering amplitude
\begin{eqnarray}
{\cal O}(f,\bar{f})&\approx&{\cal O}(\hat{q}).
\end{eqnarray}
From another point of view~\cite{Chao:1995cz}, this treatment can be connected with the on-shell condition, which maintains the gauge invariance of the hard-scattering amplitude.

Then the decay amplitude of $V \rightarrow ggg$ can be rewritten as
\begin{eqnarray}
{\mathcal A}&=&-i\sqrt{N_{c}}\int\frac{\mathrm{d}^{3}\hat{q}}{(2\pi)^{3}} f(\hat{q}) \textrm{Tr}\left[{\cal P}(\hat{q}){\cal O}(\hat{q})\right].
\end{eqnarray}
Here the hard-scattering amplitude ${\cal O}(\hat{q})$ reads
\begin{eqnarray}
{\cal O}(\hat{q}) & = & \frac{-ig^{3}_{s}}{4 N_{c}}
\bigg{(} \textrm{Tr} [T^{a_{1}}T^{a_{2}}T^{a_{3}}]
I(k_{1},k_{2},k_{3},\epsilon_{1},\epsilon_{2},\epsilon_{3})
+ \textrm{Tr} [T^{a_{2}}T^{a_{1}}T^{a_{3}}]
I(k_{2},k_{1},k_{3},\epsilon_{2},\epsilon_{1},\epsilon_{3}) \nonumber\\
& + & \textrm{Tr} [T^{a_{2}}T^{a_{3}}T^{a_{1}}]
I(k_{2},k_{3},k_{1},\epsilon_{2},\epsilon_{3},\epsilon_{1})
+ \textrm{Tr}[T^{a_{3}}T^{a_{2}}T^{a_{1}}] I(k_{3},k_{2},k_{1},\epsilon_{3},\epsilon_{2},\epsilon_{1})\nonumber\\
& + &\textrm{Tr}[T^{a_{3}}T^{a_{1}}T^{a_{2}}]
I(k_{3},k_{1},k_{2},\epsilon_{3},\epsilon_{1},\epsilon_{2})
+ \textrm{Tr}[T^{a_{1}}T^{a_{3}}T^{a_{2}}]
I(k_{1},k_{3},k_{2},\epsilon_{1},\epsilon_{3},\epsilon_{2}) \bigg{)}
\end{eqnarray}
with
\begin{eqnarray}
I&&(k_{1},k_{2},k_{3},\epsilon_{1},\epsilon_{2},\epsilon_{3}) \nonumber\\
& & = \frac{ \slashed{\epsilon}^{*}_{1} \cdot (\slashed{k}_1 - \slashed{k}_2 - \slashed{k}_3 + 2 \hat{\slashed{q}} + M )\cdot \slashed{\epsilon}^{*}_{2} \cdot( \slashed{k}_1 + \slashed{k}_2 - \slashed{k}_3 + 2 \hat{\slashed{q}} + M )\cdot \slashed{\epsilon}^{*}_{3}}{[ -k_1 \cdot k_2 - k_1 \cdot k_3 + 2(k_1 \cdot \hat{q}) + \hat{q}^2 + i \varepsilon][-k_1 \cdot k_3 - k_2 \cdot k_3 - 2(k_3 \cdot \hat{q}) + \hat{q}^2 + i \varepsilon]} ,
\end{eqnarray}
where $k_{1}$, $k_{2}$, $k_{3}$ and $\epsilon_{1}$, $\epsilon_{2}$, $\epsilon_{3}$ denote the momenta and polarization vectors of the gluons. The trace of the color generators satisfies $\textrm{Tr}[T^{a_{1}}T^{a_{2}}T^{a_{3}}]=(d^{a_{1}a_{2}a_{3}}+i f^{a_{1}a_{2}a_{3}})/4$.  Consequently, the decay amplitude consists of two parts: one proportional to the symmetric constant $d^{a_{1}a_{2}a_{3}}$ and the other proportional to the antisymmetric constant $f^{a_{1}a_{2}a_{3}}$. We show that the antisymmetric part vanishes exactly upon integration over the internal momentum $\hat{q}$. This is consistent with the numerical results in Ref.~\cite{Fu:2010gd}, which found the residual antisymmetric contribution to be negligible compared to the dominant symmetric one. So the hard-scattering amplitude ${\cal O}(\hat{q})$ reads
\begin{eqnarray}
{\cal O}(\hat{q}) & = & \frac{-ig^{3}_{s}d^{a_{1}a_{2}a_{3}}}{16 N_{c}}
\bigg{(} I(k_{1},k_{2},k_{3},\epsilon_{1},\epsilon_{2},\epsilon_{3})
+ \textrm{ 5 permutations of 1, 2 and 3.}  \bigg{)}.
\end{eqnarray}

The decay width is given by
\begin{eqnarray}\label{gamma}
\Gamma^{ggg}& = & \frac{1}{2 M}\int \frac{1}{3} \frac{1}{3!} \sum_{\mathrm{color, pol.}} \mid {\mathcal A} \mid^{2} \mathrm{d}\Phi_{3},
\end{eqnarray}
where $\int \mathrm{d}\Phi_{3}$ denotes the integration over the three-body phase space, the factor $1/3$ averages over the spins of the initial quarkonium, the factor $1/3!$ accounts for the identity of the final-state gluons, and $\sum_{\mathrm{color, pol.}}$ denotes summing over the colors and polarizations of all external particles.

For the decay $V \to ggg$, the kinematics allows one of the final-state gluons to have zero energy. Consequently, the calculation of the decay width potentially encounters an infrared (IR) divergence. The integral over $\hat{q}$ will diverge if the full $\hat{q}$-dependence in the hard-scattering amplitude is retained. Specifically, we find that an IR divergence first appears at order $\hat{q}^{4}$ (see Ref.~\cite{Bodwin:2002cfe} for more details). In essence, this occurs because the soft gluon changes the incoming $S$-wave color-singlet state into a $P$-wave color-octet state. Fortunately, according to the NRQCD factorization formalism, this IR divergence can be absorbed into the matrix elements of the $P$-wave color-octet operators~\cite{Bodwin:2002cfe}. However, these higher-order color-octet matrix elements are poorly known. By a crude estimate, Ref.~\cite{Bodwin:2002cfe} pointed out that the ${\cal O}(v^4)$ corrections were much smaller than the ${\cal O}(v^2)$ corrections. From a pragmatic perspective, and in order to present a more robust prediction, this work will focus on analytically calculating the first-order relativistic corrections.

To the first-order relativistic corrections, we get the modulus squared of the decay amplitude
\begin{eqnarray}\label{squaredamplitude}
\mid {\mathcal A} \mid^{2} & = & \left| \sqrt{N_{c}} \int\frac{\mathrm{d}^{3}\hat{q}}{(2\pi)^{3}} f(\hat{q}) {\cal M}^{0}(0) \right|^{2} \nonumber\\
& + & 2 \textrm{Re} \left[ N_{c} \int \frac{\mathrm{d}^{3}\hat{q}}{(2\pi)^{3}} \frac{\mathrm{d}^{3} \hat{q}^{\prime}}{(2\pi)^{3}} f(\hat{q}) f^{*}(\hat{q}^{\prime})  {\cal M}^{0*}(0)  {\cal M}^{1}(0) \right]
\end{eqnarray}
with
\begin{eqnarray}
{\cal M}^{0}(0) & = &\textrm{Tr}\left[ {\cal P}(0){\cal O}(0)\right]
\end{eqnarray}
and
\begin{eqnarray}
{\cal M}^{1}(0) & = & \frac{\hat{q}^{\alpha}\hat{q}^{\beta} }{2!} \left[ \left. \frac{\partial \textrm{Tr} \left[{\cal P}(\hat{q}){\cal O}(\hat{q})\right]}{\partial \hat{q}^{\alpha} \partial \hat{q}^{\beta}} \right|_{\hat{q}=0}\right].
\end{eqnarray}
Here we expand the modulus squared $\mid {\mathcal A} \mid^{2}$ to quadratic order in the internal momentum $\hat{q}$, and make the following substitution to project out the $S$-wave state:
\begin{eqnarray}
\hat{q}^{\alpha}\hat{q}^{\beta} & \rightarrow & \frac{\mathbf{\hat{q}}^{2}}{3}\left( -g^{\alpha\beta} + \frac{K^{\alpha}K^{\beta}}{M^{2}} \right).
\end{eqnarray}
Consequently, the term ${\cal M}^{1}(0) $ can be  rewritten as
\begin{eqnarray}
{\cal M}^{1}(0) & = &\frac{1}{2!} \frac{\mathbf{\hat{q}}^{2} }{3}  \left.\left[ \left( -g^{\alpha\beta} + \frac{K^{\alpha}K^{\beta}}{M^{2}} \right) \frac{\partial \textrm{Tr} \left[{\cal P}(\hat{q}){\cal O}(\hat{q}) \right]}{\partial \hat{q}^{\alpha} \partial \hat{q}^{\beta}} \right|_{\hat{q}=0} \right].
\end{eqnarray}

To proceed, we perform the integral over the internal momentum $\hat{q}$ in Eq.~\eqref{squaredamplitude}, and obtain the decay width
\begin{eqnarray}
\Gamma^{ggg}& = & \frac{160 \alpha_s^3 N_V^2 \beta_V^3}{81 \pi^{9/2} M^3}\int_{0}^{1}\mathrm{d} x_1 \int_{1-x_1}^{1} \mathrm{d} x_2 \frac{1}{x_1^3 x_2^3 (x_1 + x_2 - 2)^3}
 \left[ M^2 x_1 x_2 (x_1 + x_2 - 2) \right. \nonumber \\
&& \times \left( x_1^4 + 2 x_1^3 x_2 - 4 x_1^3 + 3 x_1^2 x_2^2 - 9 x_1^2 x_2 + 7 x_1^2 + 2 x_1 x_2^3 - 9 x_1 x_2^2 + 13 x_1 x_2 \right. \nonumber \\
&& \quad \left. \left. - 6 x_1 + x_2^4 - 4 x_2^3 + 7 x_2^2 - 6 x_2 + 2 \right) \right. \left. + 4 \beta_V^2 \left( x_1^6 + 3 x_1^5 x_2 - 6 x_1^5 + 9 x_1^4 x_2^2 \right. \right. \nonumber \\
&& \quad \left. \left. - 27 x_1^4 x_2 + 25 x_1^4 + 13 x_1^3 x_2^3 - 60 x_1^3 x_2^2 + 98 x_1^3 x_2 - 60 x_1^3 + 9 x_1^2 x_2^4 - 60 x_1^2 x_2^3 \right. \right. \nonumber \\
&& \quad \left. \left. + 153 x_1^2 x_2^2 - 177 x_1^2 x_2 + 79 x_1^2  + 3 x_1 x_2^5 - 27 x_1 x_2^4 + 98 x_1 x_2^3 - 177 x_1 x_2^2 \right. \right. \nonumber \\
&& \quad \left. \left. + 157 x_1 x_2 - 54 x_1  + x_2^6 - 6 x_2^5 + 25 x_2^4 - 60 x_2^3 + 79 x_2^2 - 54 x_2 + 15 \right) \right],
\end{eqnarray}
where $x_{i} \equiv 2E_{i}/M$ (with $E_{i}=\mid \mathbf{k}_{i} \mid$, $i=1,2$) represents the dimensionless energy fraction of the $i$-th gluon.

To discuss the symmetry in the decay process $V \rightarrow ggg$, the corresponding polarized decay widths are given by
\begin{eqnarray}\label{polarized widths}
\Gamma_{[\lambda_V, \lambda_1, \lambda_2, \lambda_3]}^{ggg}& = & \frac{1}{2 M}\int \frac{1}{3!} \sum_{\mathrm{color}} \mid {\mathcal A} \mid^{2} \mathrm{d}\Phi_{3}.
\end{eqnarray}
Here, $\lambda_V$ denotes the helicity of the heavy quarkonium, while $\lambda_i$ ($i=1,2,3$) represents the gluon helicity. The following conventions are used
\begin{eqnarray}
K & = & M (1, 0, 0, 0), \quad \varepsilon(\pm 1) = \frac{1}{\sqrt{2}}(0, \mp 1, - i, 0), \quad \varepsilon(0) = (0, 0, 0, 1)  \nonumber \\
k_{1} & = & \frac{M x_{1}}{2}(1, 0, 0, 1), \quad \epsilon_{1}(\pm 1) = \frac{1}{\sqrt{2}}(0, \mp 1, - i, 0),  \nonumber \\
k_{2} & = & \frac{M x_{2}}{2} (1, \sin\theta_{12}, 0, \cos\theta_{12}), \quad \epsilon_{2}(\pm 1) = \frac{1}{\sqrt{2}}(0, \mp \cos\theta_{12}, - i, \pm \sin\theta_{12}), \nonumber \\
k_{3} & = & \frac{M x_{3}}{2} (1, -\sin\theta_{13}, 0, \cos\theta_{13}), \quad  \epsilon_{3}(\pm 1) = \frac{1}{\sqrt{2}}(0, \mp \cos\theta_{13}, - i, \mp \sin\theta_{13}),
\end{eqnarray}
where $\theta_{ij}$ is the angle between gluons $i$ and $j$.

\section{RESULTS AND DISCUSSIONS}
\label{sec:Results and discussions}
\subsection{Polarized decay widths}
\label{subsec:Polarized decay widths}

We show a brief analysis of the symmetries in the polarized decay widths. Owing to helicity flip symmetry and phase space symmetry, the following relations hold:
\begin{eqnarray}
\Gamma_{[\lambda_V,\lambda_1,\lambda_2,\lambda_3]}^{ggg} = \Gamma_{[-\lambda_V,-\lambda_1,-\lambda_2,-\lambda_3]}^{ggg}, \quad  \quad
\Gamma_{[\lambda_V,\lambda_1,\lambda_2,\lambda_3]}^{ggg} = \Gamma_{[\lambda_V,\lambda_1,\lambda_3,\lambda_2]}^{ggg}.
\end{eqnarray}
The first relation is a consequence of helicity flip symmetry in QCD, which mandates that the decay width remains invariant under the simultaneous flip of all helicities. The second relation originates from the indistinguishability of the final-state gluons. It is worth noting that we fix the momentum of gluon $1$ along the $z$-axis, so the phase space integration is symmetric under the exchange of gluon $2$ and gluon $3$. This symmetry directly leads to the second equality. In addition, the special role assigned to gluon $1$ breaks the full permutation symmetry among the three gluons, which is why a relation symmetric in all three gluons is not manifest.

Furthermore, when summing over the polarizations of the initial quarkonium $V$, we can obtain
\begin{eqnarray}
\sum_{\mathrm{\lambda_V=-1,0,1}}\Gamma_{[\lambda_V,1,1,-1]}^{ggg}
&=&\sum_{\mathrm{\lambda_V=-1,0,1}}\Gamma_{[\lambda_V,1,-1,1]}^{ggg}
=\sum_{\mathrm{\lambda_V=-1,0,1}}\Gamma_{[\lambda_V,-1,1,1]}^{ggg}\nonumber \\
&=&\sum_{\mathrm{\lambda_V=-1,0,1}}\Gamma_{[\lambda_V,-1,1,-1]}^{ggg}
=\sum_{\mathrm{\lambda_V=-1,0,1}}\Gamma_{[\lambda_V,1,-1,-1]}^{ggg}
=\sum_{\mathrm{\lambda_V=-1,0,1}}\Gamma_{[\lambda_V,-1,-1,1]}^{ggg}, \nonumber \\
\sum_{\mathrm{\lambda_V=-1,0,1}}\Gamma_{[\lambda_V,1,1,1]}^{ggg}
&=&\sum_{\mathrm{\lambda_V=-1,0,1}}\Gamma_{[\lambda_V,-1,-1,-1]}^{ggg},
\end{eqnarray}
due to the indistinguishability of the final-state gluons and the helicity flip symmetry. The summation over the initial polarization $\lambda_V$ averages out any residual dependence on the special choice of the momentum of gluon $1$, thereby restoring the complete symmetry among all three gluons that is inherent in the physical process itself. It is noteworthy that these relations are model-independent, deriving directly from the fundamental symmetries of QCD and the indistinguishable nature of gluons. They not only provide a crucial self-consistency check for theoretical calculations, but also reveal universal features of the underlying dynamics in the decay process.

Performing analytical integration over the internal momentum of the heavy quarkonium and the three-body phase space, we obtain the polarized decay widths including the first-order relativistic corrections. As required by helicity selection rules, the following widths vanish
\begin{eqnarray}
\Gamma_{[1, 1, 1, 1]}^{ggg} =&& \Gamma_{[-1, -1, -1, -1]}^{ggg} = 0, \quad \quad \Gamma_{[1, -1, -1, -1]}^{ggg} = \Gamma_{[-1, 1, 1, 1]}^{ggg} = 0, \quad \quad \Gamma_{[0, 1, 1, 1]}^{ggg} = \Gamma_{[0, -1, -1, -1]}^{ggg} = 0,\nonumber \\
&&\Gamma_{[1, -1, 1, 1]}^{ggg} = \Gamma_{[-1, 1, -1, -1]}^{ggg} = 0, \quad \quad \Gamma_{[0, 1, -1, -1]}^{ggg} = \Gamma_{[0, -1, 1, 1]}^{ggg} = 0.
\end{eqnarray}
The nonvanishing polarized decay widths are given by
\begin{eqnarray}
\Gamma_{[1, 1, 1, -1]}^{ggg} & = & \Gamma_{[-1, -1, -1, 1]}^{ggg} = \Gamma_{[-1, -1, 1, -1]}^{ggg} = \Gamma_{[1, 1, -1, 1]}^{ggg} \nonumber \\
& = & \frac{5 \alpha_s^3 N_V^2 \beta_V^3 (3 \pi^2 - 16)}{324 M \pi^{9/2}} \left(1 - \frac{3 (304 + 39 \pi^2)}{2 (3 \pi^2 - 16)} \frac{\beta_V^2}{M^2}\right),
\end{eqnarray}
\begin{eqnarray}
\Gamma_{[1, -1, 1, -1]}^{ggg} & = & \Gamma_{[-1, 1, -1, 1]}^{ggg} = \Gamma_{[-1, 1, 1, -1]}^{ggg} = \Gamma_{[1, -1, -1, 1]}^{ggg} \nonumber \\
& = & \frac{5 \alpha_s^3 N_V^2 \beta_V^3 (29 \pi^2 - 280)}{324 M \pi^{9/2}} \left(1 - \frac{528 + 51 \pi^2}{2 (29 \pi^2 - 280)} \frac{\beta_V^2}{M^2}\right),
\end{eqnarray}
\begin{eqnarray}
\Gamma_{[0, 1, 1, -1]}^{ggg} & = & \Gamma_{[0, -1, -1, 1]}^{ggg} = \Gamma_{[0, -1, 1, -1]}^{ggg} = \Gamma_{[0, 1, -1, 1]}^{ggg} \nonumber \\
& = & \frac{10 \alpha_s^3 N_V^2 \beta_V^3}{81 M \pi^{9/2}} \left(1 - \frac{3 (11 \pi^2 - 8)}{4} \frac{\beta_V^2}{M^2}\right),
\end{eqnarray}
\begin{eqnarray}
\Gamma_{[1, 1, -1, -1]}^{ggg} & = & \Gamma_{[-1, -1, 1, 1]}^{ggg} \nonumber \\
& = & \frac{40 \alpha_s^3 N_V^2 \beta_V^3 (\pi^2 - 9)}{81 M \pi^{9/2}} \left(1 - \frac{336 + 75 \pi^2}{16 (\pi^2 - 9)} \frac{\beta_V^2}{M^2}\right).
\end{eqnarray}
Our analytical results are in full agreement with the symmetry analysis presented above. These results provide the first complete analysis of polarized decay widths for $V\to ggg$ including first-order relativistic corrections. Our calculations demonstrate how the fundamental symmetries of QCD manifest in the polarized decay patterns of the heavy quarkonium, and offer a guidance for experimentally measuring correlations among the gluon helicities.

\subsection{Unpolarized decay widths}
\label{subsec:Decay widths}

The analytical result for the unpolarized decay width $\Gamma(V \to ggg)$ including relativistic corrections is
\begin{eqnarray}
\Gamma (V \rightarrow ggg)= \frac{80(\pi^{2}-9)\alpha_{s}^{3}N_{V}^{2}\beta_{V}^{3}}{81\pi^{9/2} M } \left( 1 - \kappa \frac{\beta_{V}^{2}}{M^{2}} \right)
\end{eqnarray}
with $\kappa\equiv\frac{3(112+25\pi^{2})}{16(\pi^{2}-9)}$. This result clearly shows that the relativistic correction ($\kappa \frac{\beta_{V}^{2}}{M^{2}}\approx 0.77$) in the $J/\psi$ channel is more significant than that ($\kappa \frac{\beta_{V}^{2}}{M^{2}}\approx 0.31$) in the $\Upsilon$ channel, due to the lighter mass of charm quark. It is noteworthy that the relativistic correction in our result is smaller for a reasonable range of non-perturbative parameters than that reported in Ref.~\cite{Sang:2020zdv}, which yields an unphysical negative decay width for the $J/\psi$.

Since relativistic corrections from the internal motion of the bound quarks and QCD radiative corrections from additional gluon emissions during the hard process arise from distinct physical mechanisms, one can assume that the radiative and relativistic corrections are factorizable. Including these two corrections, we obtain the decay width
\begin{eqnarray}
\Gamma (V \rightarrow ggg) = \frac{80(\pi^{2}-9)\alpha_{s}^{3}N_{V}^{2}\beta_{V}^{3}}{81\pi^{9/2} M } \left( 1 - \kappa \frac{\beta_{V}^{2}}{M^{2}} \right) \left( 1 - C_{V} \frac{\alpha_{s}}{\pi} \right)
\end{eqnarray}
with $C_{V} = 3.7 (4.9)$ for $J/\psi (\Upsilon)$~\cite{Kwong:1987ak}.

In addition, we also compute the decay width for $V \rightarrow e^{+}e^{-}$
\begin{eqnarray}
\Gamma (V \rightarrow e^{+}e^{-}) = \frac{8 e_{Q}^{2} \alpha^2 N_V^2 \beta_V^3}{ \pi^{7/2} M} \left(1 - \frac{16 \alpha_s}{3\pi}\right),
\end{eqnarray}
where $e_{Q}$ is the charge of the heavy quark. In order to cut down the theoretical uncertainty to a large extent, we form the ratio of the gluonic to leptonic widths
\begin{eqnarray}\label{Rv}
R_{V} = \frac{\Gamma(V \rightarrow ggg)}{\Gamma(V \rightarrow e^{+}e^{-})} = \frac{10(\pi^{2}-9)\alpha_{s}^{3}}{81\pi e_{Q}^{2}\alpha^{2}}\left( 1 - \kappa \frac{\beta_{V}^{2}}{M^{2}} \right) \left( 1+C\frac{\alpha_{s}}{\pi} \right)
\end{eqnarray}
with $C=1.63$ for $J/\psi$, $0.43$ for $\Upsilon$. The common non-perturbative factor $N_V^2\beta_V^3$ cancels exactly in $R_V$, thereby yielding a more reliable prediction.

In the following numerical calculations, the values of the involved meson masses and full widths are taken from the PDG~\cite{ParticleDataGroup:2024cfk}. The QCD running coupling constants are adopted $\alpha_{s}(M_{J/\psi}/2)=0.30$ and $\alpha_{s}(M_{\Upsilon}/2)=0.21$ , which are calculated through the one-loop renormalization group equation. Using these parameters, we present the predictions for the branching ratios $\mathcal{B}(V \rightarrow ggg)$ and $\mathcal{B}(V \rightarrow e^{+}e^{-})$ as well as their ratio $R_{V}$ in Table~\ref{tab:predictions}. Here, we do not present the theoretical uncertainties, which come mainly from the QCD running coupling constant $\alpha_{s}$, and they are expected to be smaller than $50 \%$. The second column gives the results at the lowest order; the third column gives the total results including both relativistic and QCD radiative corrections. A significant deviation exists between the lowest-order predictions and the experimental data. After including both relativistic and QCD radiative corrections, our predictions show good agreement with the experimental data. Our results imply that ${\cal O}(\hat{q}^{4})$-order contributions are likely to be small in the decay processes $J/\psi (\Upsilon) \rightarrow ggg$, which is consistent with the discussion in Refs.~\cite{Chao:1995cz, Bodwin:2002cfe}.
\begin{table}[!htbp]
  \caption{\label{tab:predictions}The branching ratios $\mathcal{B}(V \rightarrow ggg (e^{+}e^{-}))$ and their ratio $R_{V}$.}
  \vspace{0.2cm}
  \centering
  \begin{tabular}{lccc}
  \hline\hline
  ~~&~~Lowest~~~ &~~Total~~&~~Exp.~\cite{CLEO:2005mdr, CLEO:2008gct, ParticleDataGroup:2024cfk}~~~  \\
  \hline
 $\mathcal{B}(J/\psi \rightarrow ggg)$~~&~~$ 453.3 \% $~~&~~$ 66.0 \% $~~&~~$(64.1 \pm 1.0 ) \%$ \\
 $\mathcal{B}(J/\psi \rightarrow e^{+}e^{-})$~~&~~$ 11.6 \% $
 ~~&~~$ 5.7 \% $~~&~~$(5.971 \pm 0.032) \% $\\
 $R_{J/\psi}$                ~~&~~$ 39.1 $~~&~~$ 11.6 $ ~~&~~$ 10.7 \pm 0.2 $  \\
  \hline
 $\mathcal{B}(\Upsilon \rightarrow ggg)$~~&~~$ 175.3 \% $~~&~~$ 81.2 \% $~~&~~$(81.7 \pm 0.7) \% $ \\
 $\mathcal{B}(\Upsilon \rightarrow e^{+}e^{-})$~~&~~$ 3.3 \% $
 ~~&~~$ 2.1 \% $~~&~~$(2.39 \pm 0.08) \% $\\
 $R_{\Upsilon}$                ~~&~~$ 53.5 $~~&~~$ 38.5 $ ~~&~~$34.2 \pm 1.2 $ \\
  \hline\hline
  \end{tabular}
\end{table}

For a cross-check, we extract the QCD running coupling constant $\alpha_{s}$ within the B-S formalism. Using our theoretical prediction for the ratio $R_{V}$ and the corresponding experimental data, we obtain the following values:
\begin{eqnarray}
\alpha_{s}(M_{J/\psi}/2) = 0.31, \quad  \alpha_{s}(M_{\Upsilon}/2)=0.20,
\end{eqnarray}
which are consistent with the values obtained from other decay channels~\cite{Bethke:2006ac}. Generally, the prediction for $\alpha_{s}(M_{\Upsilon}/2)$ is more reliable than that for $\alpha_{s}(M_{J/\psi}/2)$, because the smaller value of $\langle v^{2} \rangle_{\Upsilon}$ results in smaller higher-order relativistic corrections.

\section{SUMMARY}
\label{sec:summary}

In this paper, we investigate the decays of heavy quarkonia, $J/\psi$ and $\Upsilon$, into three gluons within the B-S formalism. The internal momentum of the quarkonium is explicitly retained in both the soft wave function and the hard-scattering amplitude. The B-S wave function is obtained by solving the B-S equation analytically, incorporating both the confinement and the one-gluon-exchange interactions. By the symmetries of the system, we first derive model-independent identities that relate different polarized decay widths, and find that all nonvanishing polarized decay widths can be categorized into four groups. For each group, we present a compact analytical expression. These results provide valuable insights for the gluon polarization.

Based on the factorization assumption, we compute the decay widths $\Gamma(J/\psi \to ggg)$ and $\Gamma(\Upsilon \to ggg)$, including both relativistic and QCD radiative corrections. In addition, we also calculate the leptonic decay widths $\Gamma(J/\psi \rightarrow e^{+}e^{-})$ and $\Gamma(\Upsilon \rightarrow e^{+}e^{-})$, and present the ratios $R_{J/\psi}$ and $R_{\Upsilon}$. Our predictions for the branching ratios $\mathcal{B}(J/\psi (\Upsilon) \rightarrow ggg)$ and  $\mathcal{B}(J/\psi (\Upsilon) \rightarrow e^{+}e^{-})$, as well as the ratios $R_{J/\psi}$ and $R_{\Upsilon}$, are in good agreement with the experimental data.  As a cross-check, we use our theoretical predictions of the ratios $R_{J/\psi}$ and $R_{\Upsilon}$ to extract the QCD running coupling constants $\alpha_s(M_{J/\psi}/2)$ and $\alpha_s(M_{\Upsilon}/2)$, which are consistent with other determinations~\cite{Bethke:2006ac}.

\section*{ACKNOWLEDGMENTS}
The work of C.-J. F. is supported by the Hubei Provincial Natural Science Foundation Youth Project under Grant No. 2024AFB151. The work of J.-K. H. is supported by the National Natural Science Foundation of China under Grant No. 12305086, the Open Fund of the Key Laboratory of Quark and Lepton Physics (MOE) under Grant No. QLPL2024P01, and the Project of Science and Technology Research Program of the Hubei Provincial Department of Education under Grant No. Q20222504. The work of C. K. is supported by the Hubei Provincial Natural Science Foundation Innovation and Development Joint Fund project under Grant No. 2024AFD010.

\newpage

\bibliographystyle{JHEP}
\bibliography{hejk}

\end{document}